\documentclass[12pt]{article}
\usepackage{amsmath}
\usepackage{amssymb}

\newcommand{\R}{\mathbb{R}}
\newcommand{\N}{\mathbb{N}}

\title{
A complex periodic QES potential and exceptional points}
\author{B Bagchi$^1$, C Quesne$^{2}$, R Roychoudhury$^3$\\ 
{\small
$^1$ Department of Applied Mathematics, University of Calcutta,} \\ {\small 92 Acharya Prafulla Chandra Road, Kolkata 700 009, India}\\ 
{\small $^2$ Physique Nucl\'eaireTh\'eorique et Physique Math\'ematique,  Universit\'e Libre de Bruxelles,} 
\\ 
{\small Campus de la Plaine CP229, Boulevard~du Triomphe, B-1050
Brussels, Belgium}\\
{\small $^3$ Physics and Applied Mathematics Unit, Indian Statistical Institute, Kolkata
700 035, India}\\ 
{\small E-mail: bbagchi123@rediffmail.com, cquesne@ulb.ac.be and rajdaju@rediffmail.com}}
\date{ }
\begin{document}
\baselineskip=22pt plus 1pt minus 1pt
%%%%%%%%%%%%%%%%%%%%%%%%%%%%%%%%%%%%%%%%%%%%%%%%%%%%%%%%%%
\maketitle

\begin{abstract} 
We show that the complex $\cal PT$-symmetric periodic potential $V(x) = - ({\rm i} \xi \sin 2x + N)^2$, where $\xi$ is real and $N$ is a positive integer, is quasi-exactly solvable. For odd values of $N \ge 3$, it may lead to exceptional points depending upon the strength of the coupling parameter $\xi$. The corresponding Schr\"odinger equation is also shown to go over to the Mathieu equation asymptotically. The limiting value of the exceptional points derived in our scheme is consistent with known branch-point singularities of Mathieu equation.
\end{abstract}

\noindent
Short title: Complex periodic QES potential

\noindent
Keywords: Quasi-exact solvability; PT symmetry; Periodic potentials; Exceptional points

\noindent
PACS number: 03.65.-w
%
%========================================================================
%
\newpage
In recent times, considerable attention has been paid to the development (see, for example, the comprehensive review \cite{bender07}) of complex non-Hermitian systems admitting space-time reflection ($\cal PT$) symmetry, the primary reason being rooted to the conjecture that the eigenvalues of the governing Hamiltonians support \cite{bender98} a real bound-state spectrum. However, the package of $\cal PT$-symmetric theories is such that should a spontaneous breaking take place, a loss of the reality of the energies is incurred and complex eigenvalues appear in conjugate pairs. In such a situation, the accompanying eigenfunctions cease to belong to the $\cal PT$ operator. The critical points where some pairs of real eigenvalues transit to the complex plane turn out to be exceptional points, {\sl i.e.}, points of the complex parameter space where two eigenvalues coalesce \cite{kato, heiss}\footnote{It should be noted that the critical points considered here are only special cases of exceptional points because (i) we restrict the range of $\xi$ in (\ref{eq:V-trig}) to real values, while we might consider the whole complex plane, and (ii) exceptional points do not necessarily occur as branch-point singularities of a single analytic function, but might also result from the equality of two analytic functions.}. Occurrences of the latter have been widely studied in the literature (see, for example, \cite{gunther, berry, leyvraz, znojil}).\par
%
%--------------------------------------------------------------------------------------------
% 
This note investigates  a new class of quasi-exactly solvable (QES) $\cal PT$-symmetric $\pi$-periodic potentials defined by
\begin{equation}
  V(x) = - ({\rm i} \xi \sin 2x + N)^2  \label{eq:V-trig}
\end{equation}
with $\xi \in \R$ and $N$ an integer, which has applications in one-dimensional crystal problems \cite{boyd}. We shall be interested here in $\pi$-periodic or $\pi$-anti-periodic solutions of the corresponding Schr\"odinger equation, which are counterparts of the band edge wavefunctions (often called eigenfunctions) of conventional theories for real periodic potentials (see, for example, ref.\ \cite{tkachuk}).  As will be demonstrated below, for odd values of $N$ ($\ge 3$), $V(x)$ leads to complex-conjugate square-root branch-point singularities, implying the crossing of the energy levels to the complex plane and giving rise to exceptional points. Here the eigenfunctions turn out to be $\cal PT$-symmetric and $\pi$-periodic in nature. In contrast, for even $N$, such critical points are absent: there are only pairs of complex-conjugate eigenvalues with corresponding eigenfunctions that are $\pi$-anti-periodic and fail to be $\cal PT$-symmetric.\par
%
%---------------------------------------------------------------------------------------------------------
%
It should be noted that the `hyperbolic' analogue of (\ref{eq:V-trig}), namely \cite{bagchi}
\begin{equation}
  V(y) = - (\zeta \sinh 2y - {\rm i} M)^2,  \label{eq:V-hyp}
\end{equation}
where $\zeta \in \R$ and $M$ an integer, is also $\cal PT$-symmetric but always has energies which are real for all integer values of $M$ and as such no critical point is encountered. A similar conclusion holds for the $\cosh$-version of (\ref{eq:V-hyp}) \cite{khare00} (see also \cite{sree}) whenever $M$ is an even integer because the QES eigenvalues occur in complex-conjugate pairs, but it does not hold for $M$ an odd integer since then the QES eigenvalues are real for $|\zeta|$ smaller than or equal to some critical value.\par
%
%-------------------------------------------------------------------------------------------
% 
Calculations carried out by us show that for the QES eigenvalues of (\ref{eq:V-trig}), critical points emerge when the quantity $N |\xi|$ keeps to a value in the vicinity of 1.47 for $N \in \N_{\rm odd}$. This prompts us to enquire into a large $N$ behaviour of the Schr\"odinger equation having $V(x)$ as its potential. In units wherein $\hbar = 2m = 1$, the Schr\"odinger equation reads
\begin{equation}
  - \frac{d^2 \psi}{dx^2} - ({\rm i} \xi \sin 2x + N)^2 \psi = E \psi  \label{eq:SE}
\end{equation}
and it is a simple exercise to convince oneself that in the large $N$ limit, equation (\ref{eq:SE}) goes to the form
\begin{equation}
  - \frac{d^2 \psi}{dx^2} - 2{\rm i} g \sin 2x \psi = \bar{E} \psi,  \label{eq:Mathieu}
\end{equation}
where $g = N \xi \in \R$ and $\bar{E} = E + N^2$. In writing (\ref{eq:Mathieu}), we have assumed $g$ to be finite under the double limits $N \to \infty$ and $\xi \to 0$.\par
%
%---------------------------------------------------------------------------------------------------------
%
Interestingly, equation (\ref{eq:Mathieu}) is the Mathieu equation \cite{abramowitz}, in which branch-point singularities are known to occur as exemplified in the works of Blanch and Clemm \cite{blanch}, Hunter and Guerrieri \cite{hunter}, and others \cite{mulholland, bouwkamp, zamastil}. A variety of numerical methods has attempted to solve Mathieu equation with an imaginary characteristic parameter as is the case in (\ref{eq:Mathieu}). Tabular values from an old paper of Mulholland and Goldstein \cite{mulholland} reflect\footnote{after making a change of the coordinate $x \to t - \frac{\pi}{4}$ without affecting the underlying periodicity.} that the eigenvalues $\bar{E}_n$ would be both real and conjugate complex: the transition occurring at $g = g_{\rm c} = 1.4687$ for solutions having a period $\pi$. It is of utmost relevance that $g_{\rm c}$ is consistent with our calculated transitional value of $N |\xi| \sim 1.47$ (where $N$ is odd corresponding to $\pi$-periodic eigenfunctions) beyond which the energy degeneracy is lifted, the critical points effecting a phase transition.\par
%
%----------------------------------------------------------------------------------------------------------
%
In a different context, Bender {\sl et al} \cite{bender99} have made a systematic WKB analysis of the potentials of form $V(y) = {\rm i} \sin^{2n+1}(y)$ ($n=0$, 1, 2,~\ldots) to justify that these have real  bands and gaps, the wavefunctions turning out to be always periodic at the edges of each band. The presence of the coupling $g$ in the Mathieu equation (\ref{eq:Mathieu}), however, makes it of a more general character and facilitates the study of critical points.\par
%
%=============================================================
%
We now turn to (\ref{eq:SE}) and proceed to interpret it as a QES system. Typically QES problems may be generated by means of a universal enveloping algebra U(sl(2)) \cite{turbiner, ushveridze}. Here we observe that in terms of the sl(2) generators
\begin{equation}
  J_+ = z^2 \frac{d}{dz} - 2j z, \qquad J_0 = z \frac{d}{dz} - j, \qquad J_- = \frac{d}{dz},
\end{equation}
the Hamiltonian $H$ can be mapped to the following spectral problem
\begin{equation}
  H_{\rm g} = 4 J_0^2 + 2\xi J_+ + 2\xi J_- - N^2 + \xi^2  \label{eq:Hg}
\end{equation}
for $j = (N-1)/2$, where in principle both even and odd values of $N$ are allowed. To arrive at the representation (\ref{eq:Hg}) we utilized the connections
\begin{equation}
\begin{split}
  H_{\rm g} &= [\mu(z)]^{-1} H \mu(z), \\
  \mu(z) &= z^{(1-N)/2} \exp\left[\frac{1}{4}\xi \left(z - \frac{1}{z}\right)\right],  \label{eq:gauge}
\end{split}
\end{equation}
with $z = \exp(2 {\rm i} x)$, as is required to have a gauge-transformed form.\par
%
%-------------------------------------------------------------------------------------------------------------
%
The eigenvalues and eigenfunctions of $H$ can now be derived from those of (\ref{eq:V-hyp}) by means of an anti-isospectral transformation $y \to {\rm i} x$, $\zeta \to - {\rm i} \xi$, $M \to N$ \cite{khare98, krajewska}. We can check that the gauge function and the gauged Hamiltonian written down in \cite{bagchi} transform to $\mu(z)$ and $- H_{\rm g}$ as they should. Thus equations (\ref{eq:Hg}) and (\ref{eq:gauge}) are in the true spirit of quasi-exact solvability.\par
%
%----------------------------------------------------------------------------------------------------------
%
Our results for some of the eigenvalues $E$ and eigenfunctions $\psi(x)$ of $H$ are furnished below ($\sigma, \tau = \pm$):
\begin{equation}
\begin{array}{ll}
  \bullet\; N = 1{\rm :} & E = - 1 + \xi^2, \qquad \psi(x) \propto e^{\frac{\rm i}{2} \xi \sin 2x}, \\[0.3cm]
  \bullet\; N = 2{\rm :} & E_{\sigma} = - 3 + 2{\rm i} \sigma \xi + \xi^2, \qquad \psi_{\sigma}(x)    
         \propto e^{\frac{\rm i}{2}\xi \sin 2x} (\cos x - \sigma \sin x),\\[0.3cm]
  \bullet\; N = 3{\rm :} & E_0 = - 5 + \xi^2, \qquad \psi_0(x) \propto e^{\frac{\rm i}{2} \xi \sin 2x} \cos 2x,
         \\[0.3cm] 
  & E_{\sigma} = - 7 + \xi^2 - 2\sigma \sqrt{1 - 4\xi^2}, \\[0.3cm] 
  & \psi_{\sigma}(x) \propto e^{\frac{\rm i}{2}\xi \sin 2x} \left[2{\rm i} \sin 2x + \frac{1}{\xi} \left(1 + \sigma
         \sqrt{1 - 4\xi^2}\right)\right], \\[0.3cm]
  \bullet\; N = 4{\rm :} & E_{\sigma,\tau} = - 11 - 2{\rm i} \sigma \xi + \xi^2 - 4\tau \sqrt{1 + {\rm i} \sigma
         \xi - \xi^2}, \\[0.3cm]
  & \psi_{\sigma,\tau}(x) \propto e^{\frac{\rm i}{2}\xi \sin 2x} (\cos x + \sigma \sin x) \left[{\rm i} \sin 2x +
         \frac{1}{\xi} \left(1 + \tau \sqrt{1 + {\rm i} \sigma \xi - \xi^2}\right)\right].          
\end{array}  \label{eq:results}
\end{equation}
\par
%
%------------------------------------------------------------------------------------------------------------------
%
The expressions for the energy and the associated eigenfunctions for higher values of $N$ are more cumbersome: of the five solutions for $N=5$, one set of energy eigenvalues
\begin{equation}
  E_{\sigma} = - 15 + \xi^2 - 2\sigma \sqrt{9 - 4\xi^2}  \label{eq:eigenvalues}
\end{equation}
is obtained by solving a quadratic equation and the corresponding eigenfunctions are
\begin{equation}
  \psi_{\sigma}(x) \propto e^{\frac{\rm i}{2}\xi \sin 2x} \cos 2x \left[2{\rm i} \sin 2x + \frac{1}{\xi} \left(3 + 
  \sigma \sqrt{9 - 4\xi^2}\right)\right],
\end{equation}
while the other set corresponds to the cubic equation
\begin{equation}
  {\cal E}^3 + 20 {\cal E}^2 + 64 (1 + \xi^2) {\cal E} + 768 \xi^2 = 0,  \label{eq:cubic}
\end{equation}
where ${\cal E} = \xi^2 - 25 - E$, with eigenfunctions
\begin{equation}
  \psi(x) \propto e^{\frac{\rm i}{2}\xi \sin 2x} (\sin^2 2x + b \sin 2x + c).  \label{eq:psi-cubic}
\end{equation}
In (\ref{eq:psi-cubic}), the quantities $b$ and $c$ are given by the relationships
\begin{equation}
  b = - \frac{\rm i}{4\xi} (16 + {\cal E}), \qquad c = \frac{24 + {\cal E}}{\cal E}.  \label{eq:b-c}
\end{equation}
The three solutions of $\cal E$ from (\ref{eq:cubic}), when substituted in (\ref{eq:b-c}), yield the corresponding values of $b$ and $c$ which, in turn, provide the expressions for the eigenfunctions from (\ref{eq:psi-cubic}).\par
%
%----------------------------------------------------------------------------------------------------
%
We thus notice that for even $N$, $\cal PT$ symmetry is broken: only pairs of complex-conjugate eigenvalues exist, the corresponding eigenfunctions being $\pi$-anti-periodic. On the other hand, for odd $N$, there is at least one real eigenvalue irrespective of whether $\cal PT$ symmetry is broken or not, while the remaining ones are either real or in complex-conjugate pairs depending upon a critical value of $\xi$. However, the corresponding eigenfunctions are $\pi$-periodic but their being $\cal PT$-symmetric or not depends upon the nature of eigenvalues.\par
%
%-----------------------------------------------------------------------------------------------------
% 
While it is trivial to see from (\ref{eq:results}) that for $N=3$ there are critical points at $|\xi| = 0.5$, determining those for the case $N=5$ requires solving the cubic equation (\ref{eq:cubic}). A little algebra shows that the condition for the existence of one real and a pair of complex-conjugate roots relies upon the positivity of $\Delta$ that is defined by
\begin{equation}
  \Delta = 16 \xi^6 - 4 \xi^4 + 103 \xi^2 - 9.
\end{equation}
\par
%
%-----------------------------------------------------------------------------------------------------------
%
Solving for $\Delta=0$ gives the roots $\xi^2 = 0.0876$ and $0.0812 \pm {\rm i} (2.5331)$, from which we immediately identify critical points in the neighbourhood of $|\xi| = 0.296$. The other branch-point singularities at $|\xi| = 1.5$, where the eigenvalues (\ref{eq:eigenvalues}) coalesce, are situated further away from those coming from the cubic equation (\ref{eq:cubic}) and for the purpose of comparison with Mathieu equation do not interest us.\par
%
%----------------------------------------------------------------------------------------------------------
%
The above procedure can be continued further. For $N=7$, we are led to a cubic equation
\begin{equation}
  {\cal E}^3 + 56 {\cal E}^2 + 16 (49 + 4\xi^2) {\cal E} + 768 (3 + 2\xi^2) = 0
\end{equation}
and a quartic one
\begin{equation}
  {\cal E}^4 + 56 {\cal E}^3 + 16 (49 + 10 \xi^2) {\cal E}^2 + 384 (6 + 17 \xi^2) {\cal E} + 2304 \xi^2
  (24 + \xi^2) = 0,
\end{equation}
where ${\cal E} = \xi^2 - 49 - E$. A search for their zeros tells us that there are critical points coming from the latter and situated at $|\xi| = 0.2107$ further which the eigenvalues develop square-root singularities. The corresponding eigenfunctions are too complicated to be given here.\par
%
%-------------------------------------------------------------------------------------------------
% 
Putting everything together, it is significant to note the convergence of the quantity $N|\xi|$ to the value of 1.47. Indeed our findings are
\begin{equation}
\begin{array}{lll}
  & N |\xi| = 1.5 & \qquad \mbox{\rm for $N=3$} \\[0.3cm]
  & \hphantom{N |\xi|} = 1.4797 & \qquad \mbox{\rm for $N=5$} \\[0.3cm]
  & \hphantom{N |\xi|} = 1.4749 & \qquad \mbox{\rm for $N=7$} 
\end{array}
\end{equation}
which are comfortably close to Mulholland-Goldstein estimate of $g_{\rm c} = 1.4687$. The trend prevails for higher odd integers $N$ as well, as our numerical results have indicated. We therefore conclude that the QES model given by (\ref{eq:V-trig}) is well described by an imaginary sine potential in the large $N$ limit. The presence of the imaginary coupling constant ${\rm i} \xi$ makes the role of the critical points very specific as comparison with the numerical solutions of the Mathieu equation has shown.\par
%
%======================================================================
% 
Summarizing, we have proposed a QES $\cal PT$-symmetric periodic potential which possesses real and/or complex-conjugate eigenvalues. The latter may assume square-root branch-point singularities signalling the appearance of exceptional points. It is also established that our model goes over to the Mathieu equation asymptotically and that, even for rather moderate odd values of $N$, there is a remarkable coincidence of the limiting value of the lowest exceptional points found in our scheme with the estimates of the lowest branch-point singularities of Mathieu equation \cite{mulholland, bouwkamp}.\par
%
%====================================================================
%  
\section*{Acknowledgments}

We would like to thank an anonymous referee for some helpful comments and suggestions. Two of us, BB and RR, gratefully acknowledge the support of the National Fund for Scientific Research (FNRS), Belgium, and the warm hospitality at PNTPM, Universit\'e Libre de Bruxelles, where this work was carried out. One of us (RR) is also grateful to the Council of Scientific and Industrial Research (CSIR), New Delhi, for a grant (project no 21/0659/06/EMR-II).\par
%
%==========================================================
%

\end{document}